\documentclass[11pt,leqno]{article}
\usepackage{latexsym}
\usepackage{epsfig}
\usepackage{subfigure}

\newtheorem{theorem}{Theorem}[section]
\newtheorem{lemma}[theorem]{Lemma}

\newtheorem{claim}[theorem]{Claim}

\long\def\remarks #1{\noindent{\bf Remarks:} #1\\}
\newenvironment{proof}{\noindent{\bf Proof:}}{\hfill $\Box $\\}

\def\sqr#1#2{{\vcenter{\vbox{\hrule height .#2pt
              \hbox{\vrule width .#2pt height#1pt \kern#1pt
              \vrule width .#2pt} \hrule height .#2pt}}}}
\def\case #1: #2.{\bigskip \noindent {\bf Case #1:} {\it #2.}}
\def\xcase #1 #2{\noindent {\bf Case #1: #2}\ }
\long\def\claim #1: #2.{\bigskip\noindent {\bf Claim #1:} {\it #2.}}
\long\def\xclaim #1: #2.{\noindent {\bf Claim #1:} {\it #2.}}
\def\claimx #1.{\bigskip\noindent {\bf Claim:} {\it #1.}\bigskip}
\def\part #1 #2{\bigskip \noindent {\bf #1} {\it #2}}
\def\ipart #1: #2.{\bigskip \noindent {\it #1:} {\it #2.}}
 
\def\o {\overline}

\def\nclaim #1 {\noindent{\bf Claim #1: }}

\begin{document}   

\title{Set-merging for the Matching Algorithm of Micali and Vazirani}
\author{Harold N.~Gabow%
\thanks{Department of Computer Science, University of Colorado at Boulder,
Boulder, CO 80309.
E-mail: {\tt hal@cs.colorado.edu}}
}

\date{July 4, 2013}

\maketitle

%\headline={\ifnum\pageno=1{\hfill\today}\else{\hfill}\fi}
%\ifnum \pageno>1 {}\else
%
\def\today{\ifcase\month\or
January\or February\or March\or April\or May\or June\or
July\or August\or September\or October\or November\or December\fi
\ \number\day, \number\year}
\def\date#1.#2.{\ifcase#1\or
January\or February\or March\or April\or May\or June\or
July\or August\or September\or October\or November\or December\fi
\ #2, \number\year}
\def\ydate#1.#2.#3.{\ifcase#1\or
January\or February\or March\or April\or May\or June\or
July\or August\or September\or October\or November\or December\fi
\ #2, 199#3}
\def\nydate#1.#2.{\ifcase#1\or
January\or February\or March\or April\or May\or June\or
July\or August\or September\or October\or November\or December\fi
\ #2}
\def\doublespace{\multiply\baselineskip by3\divide\baselineskip by2%
                 \def\doublespace{}}%don't doublespace again!
\def\bigdoublespace{\multiply\baselineskip by2%
                 \def\bigdoublespace{}}%don't doublespace again!
\def\imp{\ifmmode {\ \Longrightarrow \ }\else{$\ \Longrightarrow \ $}\fi}
\def\rimp{\ifmmode {\ \Longleftarrow \ }\else{$\ \Longleftarrow \ $}\fi}
\def\ximp{\ifmmode {\Longrightarrow\ }\else{$\Longrightarrow\ $}\fi}
\def\xrimp{\ifmmode {\Longleftarrow\ }\else{$\Longleftarrow\ $}\fi}
\def\iff{\ifmmode {\ \Longleftrightarrow \ }\else{$\ \Longleftrightarrow \ $}\fi}
\def\xiff{\ifmmode {\Longleftrightarrow\ }\else{$\Longleftrightarrow\ $}\fi}
\def\tru{\ {\bf true}\ }
\def\fal{\ {\bf false}\ }
\def\wrt{\ {\it wrt}\ }
\def\endskip{\medskip}%\smallskip
% Nov 19 2013: i redefined \qed to a hollow box
%\def\qed{\  \vrule width4pt depth-1pt height7pt\endskip} %\bigskip }
\def\qed{$\Box$}
\def\qedn{\ \vrule width4pt depth-1pt height7pt }
\def\rqed{\hfill\hbox to 24 pt{\vrule width4pt depth-1pt
height7pt\hfil}\bigskip}
\def\rqedn{\hfill\hbox to 24 pt{\vrule width4pt depth-1pt height7pt\hfil}}
%\def\rsqed{\hfill\vrule width4pt depth-0.5pt height5.5pt\endskip}
% old version:\def\log{\ifmmode \,\hbox{log}\,\else{\it log }\fi}
% this puts 10point roman into superscritps
% new version
\def\log{\ifmmode \,{ \rm log}\,\else{\it log }\fi}
\def\con {\subseteq}
\def\pcon{\subset}
% the old step macro, \numstp (formerly \stp) gives a numbered step
\def\firstnumstp#1 {\bigskip \noindent{\it Step} #1.\newquad}
\def\numstp#1 {\endskip\noindent{\it Step} #1.\newquad}
% the new macro, \stp gives a named step
\def\newquad{\hskip1ex}
\def\stp#1.{\endskip
%\penalty-1000
\noindent{\it #1 Step.}\newquad}
\def\firststp#1.{\bigskip
\penalty-1000
\noindent{\it #1 Step.}\newquad}
\def\cas#1 {\smallskip\noindent{\bf Case} #1.\ } %\quad}
%
%the 1st sec macro allowed breaks at the end of a page
%\def\sec#1{\bigskip\bigskip\noindent{\bf #1}\bigskip}
%
%the 2nd encouraged it to go to the next page
%\def\sec#1{\bigskip
%\penalty-2000%
%\noindent{\bf #1\hfill\break}%\hskip\parindent
%\hbox to \parindent{\hfill}\ignorespaces}
%%\bigskip}
%
%the 3rd set section heading in 12 point type
\long\def\sec#1{\bigskip
\penalty-2000%
\noindent{\twelvebf #1}\par\ignorespaces\noindent\ignorespaces}
%with this sec macro for acknowledgments and bibliography
\def\aorbsec#1{\noindent{\twelvebf #1}}
\def\nsec#1{\penalty-2000%
\noindent{\bf #1\hfill\break}%\hskip\parindent
\hbox to \parindent{\hfill}\ignorespaces}
\long\def\res #1. #2{\bigskip
\penalty-1000
\noindent {\bf #1.}\newquad%
#2 \bigskip}
\long\def\nres #1. #2{\bigskip
%\penalty-1000
\noindent {\bf #1.}\newquad%
#2}
\def\pf{\noindent {\bf Proof.}\newquad}
\def\cont{\ifmmode\star\else$\star$\fi}
% {\ \ \hbox{---}\kern-9.8pt\mid\kern-6pt/\kern-5pt\backslash\ }
% \else{$\ \ \hbox{---}\kern-9.8pt\mid\kern-6pt/\kern-5pt\backslash\ $}\fi}
\def\+{\tabalign} %ordinarily \+ is \outer
\def\nskp{\def\bigskip{}}
\def\i{($i$) } \def\xi{($i$)}
\def\ii{($ii$) } \def\xii{($ii$)}
\def\iii{($iii$) } \def\xiii{($iii$)}
\def\iv{($iv$) } \def\xiv{($iv$)}
\def\pa{({\it a}) } \def\xpa{({\it a})} 
\def\pb{({\it b}) } \def\xpb{({\it b})}
\def\pc{({\it c}) } \def\xpc{({\it c})}
\def\hi{\hskip20pt\i} \def\hii{\hskip20pt\ii} \def\hiii{\hskip20pt\iii}
\def\ha{\hskip20pt\pa} \def\hb{\hskip20pt\pb} \def\hc{\hskip20pt\pc}
\def\tran{{\buildrel*\over\to}}
\def\n{\rlap{$\>/$}}
\def\({{\rm(}} \def\){{\rm)}}
\def\c#1{\lceil {#1} \rceil}
\def\f#1{\lfloor {#1} \rfloor}
\long\def\boxit#1{\vtop{\hrule
\hbox{\vrule\quad\vtop{\vskip5pt\hbox{#1}\vskip5pt}\quad\vrule}
\hrule}} %usually #1 is a \vtop
\def\iboxit#1{\vtop{\hrule
\hbox{\vrule\quad\vtop{\vskip5pt\hbox{{\it #1}}\vskip5pt}\quad\vrule}
\hrule}} %usually #1 is a \vtop
\def\x{\iffalse}
\def\b{\bigskip}
\def\set #1#2{\{ #1:#2 \}}
\def\pset #1#2{( #1:#2 )}
\def\h{\hskip20pt}
\def\hi{\advance\parindent by 20pt}

%\magnification=\magstephalf\bigdoublespace
%
\def\rqed{\hfill$\Box$}
%
%\hbox to 24 pt{\vrule width4pt depth-1ptheight7pt\hfil}\bigskip}
%
\def\rqedn{\hfill$\Box$}
%
%\hbox to 24 pt{\vrule width4pt depth-1pt height7pt\hfil}}
%
\def\o{\overline}
\def\C.{\mathy {\cal C}}
\def\E.{\mathy {\cal E}}
\def\G{\ifmmode {{\cal G}}\else{{$\cal G$}}\fi}
\def\T.{\ifmmode {{\cal T}}\else{{$\cal T$}}\fi}
\def\B.{\ifmmode {{\cal B}}\else{{$\cal B$}}\fi}
\def\D.{\ifmmode {{\cal D}}\else{{$\cal D$}}\fi}
\def\A.{\ifmmode {{\cal A}}\else{{$\cal A$}}\fi}
\def\I.{\ifmmode {{\cal I}}\else{{$\cal I$}}\fi}
\def\F.{\ifmmode {{\cal F}}\else{{$\cal F$}}\fi}
\def\S.{\ifmmode {{\cal S}}\else{{$\cal S$}}\fi}
\def\L.{\ifmmode {{\cal L}}\else{{$\cal L$}}\fi}
\def\M.{\mathy {\cal M}}
\def\X.{\ifmmode {{\cal X}}\else{{$\cal X$}}\fi}
\def\Y.{\ifmmode {{\cal Y}}\else{{$\cal Y$}}\fi}

\def\P.{\ifmmode {\Pi}\else{{$\Pi$}}\fi}

\def\mathy #1{\ifmmode {#1}\else{$#1$}\fi}
\def\bcl #1{\mathy {#1\cup B_{#1}}}
\def\ldot{.}
\def\msum#1#2#3{\bigvee_{#1}^{#2} {#3}}
\def\h{\hskip20pt}
\def\cl #1{\ifmmode {{[#1]}}\else{{$[#1]$}}\fi}
\def\fl #1{\ifmmode {{\lfloor #1\rfloor}}\else{{$\lfloor#1\rfloor$}}\fi}
\def\cll #1{\ifmmode {{[#1]_2}}\else{{$[#1]_2$}}\fi}
\def\cle #1{\ifmmode {{[#1]_e}}\else{{$[#1]_e$}}\fi}
\def\od{\overline d}
\def\lf.{{\it l\_find}}
\def\lfa #1.{{\it l\_find}$(#1)$}
\def\l #1(#2){{\it l\_find}\ifmmode_{#1}(#2)\else$_{#1}(#2)$\fi}
\def\a{({\it a}) }\def\xa{({\it a})}
\def\bb{({\it b}) }\def\xb{({\it b})}
\def\rv{\rho^{_V}}
\def\fv{\F.^{^{_V}}} 

\def\2nca{{\tt nca}}

\def\id{{\hbox{{\rm idom}}}}

\input epsf

\begin{abstract}
The algorithm of Micali and Vazirani \cite{MV} 
finds a maximum cardinality matching
in time $O(\sqrt n m)$ if an  efficient set-merging algorithm is used.
The latter is provided by the
incremental-tree set-merging algorithm of
\cite{GabTar}. Details of this application to matching 
were omitted from \cite{GabTar} and are presented in
this note.
\end{abstract}

\def\v{({\it v}) }
\def\xv{({\it v})}
\def\to{\!\!\rightarrow\!\!}

\def\fini{{\tt find}}
\def\unii{{\tt union}}
\def\ncai{\mathy {\uparrow}}
\def\spai{{\it span}}
\def\invi{\ifmmode{\uparrow^{-1}}\else{$\uparrow^{-1}$}\fi }
\def\joii{{\it c\_count}}
\def\reai{{\it closers}}

\def\finx{\fini\ }
\def\unix{\unii\ }
\def\ncax{\ncai\ }
\def\spax{\spani\ }
\def\invx{\invi\ }
\def\joix{\joii\ }
\def\reax{\reai\ }

\def\arg #1{\ifmmode {(#1)} \else $(#1)$\fi}
\def\aarg #1{\ifmmode {[#1]} \else $[#1]$\fi}
\def\mathy #1{\ifmmode {#1} \else{$#1$}\fi}

\def\fin #1{\fini \arg {#1}}
\def\uni #1{\unii \arg {#1}}
\def\nca #1{\mathy {#1}\ncai}
\def\spa #1{\spai \arg{#1}}
% arrays:
\def\inv #1{\invi \aarg{#1}}
\def\joi #1{\joii \aarg {\,[#1]\,}}
\def\rea #1{\reai \aarg {#1}}
%HERE

\def\dg{\mathy {{\cal D}(G)}}
\def\Nx {\mathy {\widehat M}}
\def\N #1{\mathy {\widehat M (#1)}}
\def\tl #1{{\it tail}\ifmmode (#1) \else{$(#1)$}\fi}
\def\hd #1{{\it head}\ifmmode (#1) \else{$(#1)$}\fi}

\def\groi{{\tt join\_closers}}
\def\grox{\groi\ }
\def\gro #1{\groi$(#1)$}
\def\v #1{{\tt visit}$(#1)$}

\def\cl #1{\ifmmode {{[#1]}}\else{{$[#1]$}}\fi}
\def\F.{\mathy {\cal F}}

% DDFS macros
\def\ha{\widehat A}
\def\hb{\widehat B}

\def\bd{\hbox{bud}}
\def\bs{\hbox{base}}
\def\pl{\hbox{petal}}
\def\pls.{\hbox{petal}$^*$}
\def\bds.{\hbox{bd}$^*$}

\def\l{{\hbox{\rm level}}}
\def\ml{{\hbox{\rm minlevel}}}
\def\Ml{{\hbox{\rm maxlevel}}}
\def\el{{\hbox{\rm evenlevel}}}
\def\ol{{\hbox{\rm oddlevel}}}
\def\t{{\hbox{\rm tenacity}}}

\section {Introduction}
Micali and Vazirani implemented the ``shortest augmenting path'' approach
to find a maximum cardinality matching in 
time $O(\sqrt n m)$ \cite{MV}.  
This time bound assumes a linear-time algorithm
for set-merging is used.
It is tempting to conjecture (as was done in \cite{MV}) that the
special structure of blossoms achieves this goal automatically. However
this claim remains unsubstantiated. Instead
the
incremental-tree set-merging algorithm of
\cite{GabTar} can be used to represent blossoms and achieve the desired time
bound. 
The details of this application to matching were omitted from \cite{GabTar}
and they are presented in
this note.
For accuracy we use the recent paper \cite{V} as the source of 
all details for the cardinality matching algorithm. We assume familiarity
with that paper.%
\footnote{Another cardinality matching algorithm running in time
$O(\sqrt n m)$ is presented in \cite{GabTarMatch}.
The use of incremental-tree set-merging in that algorithm is straightforward.}

We will present
an implementation of the Micali-Vazirani algorithm that achieves
time $O(m)$ per phase on a RAM. This easily implies the desired
time bound $O(\sqrt n m)$. The issue is the use of
set-merging to represent blossoms. Towards  this end
we divide each phase into two subphases.
The first subphase
assigns levels to vertices. In this subphase 
all free vertices are essentially equivalent. This allows the search structure
to be modeled by a tree, and allows use of 
incremental-tree set merging to model blossoms.
The second subphase  finds augmenting paths. 
Here we choose bridges for DDFS in a
depth-first manner. This gives a linear order to
newly created blossoms, allowing set merging to be handled with a stack.
(This stack-based set-merging is used in
\cite{GabSC} to compute strong components efficiently.)
The first subphase requires a RAM, the second subphase runs on 
both RAM and pointer
machine.

The rest of this section gives background from \cite{V}.
Then the set-merging algorithm is presented,
the first ``assignment'' subphase in Section
\ref{AssignmentSec} and the second 
''augmentation'' subphase in Section
\ref{AugmentationSec}.

The entire discussion uses this terminology:
A {\em vertex} is a vertex of the given graph $G$.
A {\em blossom} is a blossom from the previous search level (i.e., tenacity
$<2i+1$ during search level $i$);
a {\em base} is the base vertex of such a blossom,
or a vertex of finite \el\ and empty blossom.
Thus at the start of 
a search level, the vertices of $G$ with finite \el\  
are partitioned into 
blossoms and bases.

\iffalse
The remaining vertices with finite \ml\ 
(i.e.,  vertices with finite \ol\ not in any blossom) are called {\em inner
vertices}. (THIS DIFFERS FROM THE PAPER)
All other vertices have
infinite \ml.
\fi

As in \cite[Definition 29]{V}
every  vertex $v$ has a value $\bd^*(v)$.
A {\em petal$^*$} is the set of all vertices with
finite \el\ that have
 the same $\bd^*$.
Assume 
%(perhaps DIFFERENT FROM PAPER) 
a blossom has bud equal to
its base, so when a search level begins, each blossom plus its base
is a \pls..
At any point in time the \pls.'s partition the set of vertices 
with finite \el.
A {\em free \pls.} is a \pls. whose \bds. is a free vertex.
We say $\bd^*(P)=b$ if $P$ is a \pls. corresponding to $\bd^*\ b$.

Note that $\bd^*(v)=v$
iff $v$ is 
the bud of a \pls., or (as in \cite{V}) $v$
is not in any \pls..
% has finite \ol\ but 
%
%or $v$ has infinite \ml).

\def\pto #1{p[v \hbox{ to } #1]}

The following lemma 
is adapted from \cite{V}. 
%which may or may not be proved in the paper but I THINK IS TRUE.
We will use parts \i and \xiii;
we include \ii for completeness.

\begin{lemma}
\label{PetalLemma}
Consider an arbitrary  \pls. $P^*$ with bud $b$. 
Let $v$ be an arbitrary 
vertex in $P^*$.

\i Every $\ml(v)$ path $p$ goes through $b$.
Its subpath from $v$ to $b$, denoted
$\ml(v;b)$ (like
the notation of \cite{V})
 is contained in $P^*$.

\ii The same holds for a $\Ml(v)$ path $p$
whose (unique) $\t(v)$ bridge has been processed by MAX. 

\iii $P^*$ is precisely equal to the union of the paths
$\ml(v,b)$, $v\in P^*$. 
In fact we can restrict the union to vertices $v$ 
in the support of
$P^*$, assuming we add in the \pls. whose $\bd^*$ is $v$.
\end{lemma}

\remarks
{Note that \ii
need not hold for a \Ml\ path using
a bridge that has not been processed by MAX (i.e., the petal has
not been expanded to its complete blossom).

Also note that parts \xi--\ii imply the  Micali-Vazirani algorithm
handles augmenting paths correctly: Suppose $P^*$ of the lemma
is on an augmenting path; clearly it contains $b$. 
Suppose for contradiction that
$v\in P^*$ is on another ap 
 (ap $=$ shortest augmenting path)
$Q$ that does not contain $b$. Part \i shows 
$Q$ cannot contain a $\ml(v)$ path.
\cite[Theorem 26]{V} shows a $\Ml(v)$ path $M$ in $Q$ uses a bridge of tenacity
$<l_m$.
So part \ii applies to $M$. Thus $M$  contains $b$ and cannot be in $Q$.
}
%\b

\begin{proof}
Let $t$ be the tenacity in the current search level.

Define a subgraph $H(P^*)$: Start with the graph in which $P^*$ is formed
(i.e., previous augmenting paths, if any, have been deleted).
Then remove every bridge that has not been processed by MAX.
Apply 
\cite[Lemma 30]{V} 
to $P^*$.
If $P^*$ has no tenacity $t$ vertices the lemma shows $P^*-b$ is a blossom
of tenacity $t-2$. In the opposite case $P^*-b$ is a blossom
of $H(P^*)$ of tenacity $t$. Since $H(P^*)$ has $l_m>t$, 
\cite[Theorem 7]{V} 
applies.
It shows that as in
parts \i and \xii, each $\ml(v)$ and $\Ml(v)$ path $M$ passes through $b$..

\cite[Definition 17]{V} 
of blossom, plus a simple induction, 
%(OR SOME THEOREM PROVED IN THE PAPER?) 
show $M$ is contained in $P^*$. This completes the proof
of \i and \xii.
It also proves \xiii.
\end{proof}

\iffalse
Since the DDFS reached $v$, 
the successor of $v$ in $\pto b$ was visited. 
Applying this observation inductively shows the entire subpath $\pto b$
was visited in the DDFS and hence is in $P^*$.
\fi

\section{Assignment Subphase}
\label{AssignmentSec}
Each phase begins in the Assignment Subphase. 
This subphase assigns levels, according to the procedures MIN and MAX
of \cite[Algorithm 1]{V}.
The subphase ends when a DDFS discovers the first augmenting path.
At that point the Augmentation Subphase begins.

The Assignment Subphase is based on a search forest $F$.
Section \ref{STSec} defines $F$ and discusses 
its properties.
Then Section \ref{LASec} 
gives the algorithm for the subphase.

\iffalse
Define the index of the last search level $i_m=\f{l_m/2}$.
The following
algorithm is used for each search level
$i< i_m$.

i assumed we weren't in $l_m$. if we were, the algorithm would run along
with DDFS discovering some new petals, and eventually some DDFS would
discover an ap.
\fi

\subsection{The Search Forest}
\label{STSec}
At any point in a phase consider the following
forest $F$ (which represents the current search structure). 
$F$ has a node  for every \bds. $v$ that has 
 a finite \ml.
Each free \pls. is a root.
(At the start of the phase this means each free vertex is a root.)
The parent of $v$ corresponds to the first predecessor of $v$.
More precisely
suppose MIN scans an edge $(u,v)$  
and changes \ml$(v)$ from $\infty$ to a finite value.
Here $u$ is a vertex, and
$\bd^*(u)$ is a node of $F$.
The new tree $F$ has $\bd^*(u)$ the parent of $v$.  
Thus the $F$-path from a node $v$ 
(a $\bd^*$) to its root, say $F(v)$,
corresponds (by contraction) to a $\ml(v)$ 
path of props from $v$ to a free \pls..

This completes
the definition of $F$. Now we  describe how $F$
gets modified by
the MIN and MAX procedures.

If MIN assigns a first predecessor to
$v$, $F$ gets a  new leaf $v$.

Suppose MAX 
forms a new petal $P$
for bridge $rg$ of tenacity $t=2i+1$.
Let $P^*$ be the corresponding
\pls., and let $b$ be its bud
(i.e., the bottleneck of the DDFS for $rg$).
Take any $v\in P^*$ that is a \bds. at the start of  the current DDFS.
Lemma \ref{PetalLemma}\i
shows the $vb$-subpath of $F(v)$ 
is contained in $P^*$. 
So  Lemma \ref{PetalLemma}\iii 
shows the entire petal $P^*$ corresponds to a subtree $S$ of $F$.
We conclude the new $F$ is formed by contracting the edges
of $S$ in $F$.

\subsection{The Algorithm}
\label{LASec}
We use 
incremental-tree set merging as the data
structure for $F$. Let $D$ be the forest
of this data structure. A \bds. $b$ in $F$
corresponds to a subtree of $D$ 
that has been merged into its root
node $b$. Beyond that the two forests are identical.

The MIN step performs a {\em grow} operation
on $D$ each time a first predecessor is processed.

A MAX step 
will execute DDFS and either find the next \pls.
or find the first augmenting path. 

First suppose  DDFS, say 
for  bridge $rg$, 
identifies a \pls. $P^*$
with bud $b$.
The DDFS has identified
the support of $rg$
(in the trees $R$ and $G$ of \cite{V}).
The algorithm 
performs
a {\em union} operation on $D$ for
each support
node. This merges the entire
petal $P^*$ into its bud  $b$. 
(To verify this we remark that  a \pls. $Q^*$ whose bud$^*$ $q$ is in the support
of $rg$ is contained in $P^*$,
since every vertex
of $Q^*$ gets a new \bds. $b$.
So the {\em union}s correctly contruct $P^*$.)
The updated $D$ corresponds to the new search forest $F$ by
the description of MAX in Section
\ref{STSec}.

\iffalse
IS THERE A MINOR TERMINOLGICAL ISSUE, caused by the fact that
some tenacity $t$ vertices of $Q^*$ may be in the support of $rg$
and others NOT in the support?
\fi

\b

It remains to describe how the DDFS is implemented, and
in particular how it detects  the first augmenting path.
Assume the DDFS begins with the sets of $D$ correctly corresponding
to the current \pls.s.
The DDFS uses {\em find} operations on $D$
to jump from a vertex to its \bds. $x$; it moves from $x$ using a prop.
Observe that the above procedure to update $F$ and the data structure
$D$ occurs after the  DDFS. So 
all the  {\em find} operations in
the DDFS give the correct $\bd^*$. Thus the DDFS
works correctly, discovering the next \pls. (which
might be a free \pls.) or the first augmenting path.

\subsection{Time Bound}

The time for the Assignment Subphase is $O(m)$. In proof,
the time for a Micali-Vazirani phase is $O(m)$ plus the time for set merging.
Incremental-tree set merging uses $O(n+m)$ time for 
$m$ {\em find} operations, $n$ {\em make\_root} or {\em grow}
operations, and $n$ {\em union} operations.

\section{Augmentation Subphase}
\label{AugmentationSec}

The last search level $i_m$ of the phase is implemented by
the Augmentation Subphase.
During this subphase
the algorithm will know $\bd^*(v)$ for every vertex $v$ in the following sense. At the start of the phase $\bd^*(v)=\bs^*(v)$ from phase $i_{m-1}$. 
At some point
$v$ may enter a search stack. It remains in the stack until it is deleted,
either because of an ap or because it is  known not to be in any ap.
From this point on 
$\bd^*(v)$ will always be computed using the data structure 
for the stack.
 
The stack  models this variant of DDFS: 
A {\em DDFS anchored on $A$}, for bridge $rg$, 
starts with a (shortest) path $A$ from level $l_0$ to a vertex $r$ 
which is on a bridge $rg$ of tenacity $l_m$.
Denote $A$ as the sequence $A_1,A_2,\ldots, A_\ell$, where each $A_i$ is a vertex
or a $\pl^*$ (a vertex $A_i$ will be a vertex of finite \ol\ not in any \pls.).
As usual the  DDFS 
visits some set of vertices,
say $X$, 
and finds either an ap or a new $\pl^*$ which actually is $X$.
The anchored DDFS has this defining property:
If an ap is  found then $V(A)\con X$.
If $X$ is a $\pl^*$, say with bud $b$, then $b=\bd^*(A_i)$ for some $i$,
and $V(A)\cap X $ consists of the sets $A_j, j\ge i$.

We note that the algorithm will be  valid for a pointer machine.

\subsection{Data Structure}

Array $\ha$ stores the sequence $V(A_1),\ldots, V(A_\ell)$.
Each vertex $v$ %(i.e., vertex of given graph $G$)
%\em of $G$} the given graph (or the graph with blossoms 
%from penultimate search shrunk) 
has a pointer (index)
\[ %\begin{equation}
a[v]=\left\{
\begin{array}{cc}
0 & v\notin V(A)\\
i&\ha[i]=v.
\end{array}
\right.
\]%\end{equation}
A companion array gives the buds for $A$: 
\[\hb[i] = \bd^*(A_i),\ i=1,\ldots, \ell.\]
$\ha$ stores
$\bd^*(A_i)$ before the other vertices of $A_i$.
Thus for a vertex of $G$ in $V(A)$, $\bd^*(v)$ is the vertex $b=\hb[i]$
where $i$ is the maximum index satisfying $a[b]\le a[v]$. 

In addition there are some bookkeeping lists, that are easily
maintained: Each $\pl^*$ $P$ has a list of props $uv$, $u\in P$ and
$u$ a predecessor of $v$, and bridges $uv$ of tenacity $\ell_m$, $u
\in P$ ($v$ may or may not be in $P$).
Each blossom (from $i_{m-1}$) has a list of its $G$-vertices
(it will be used to populate $\ha$).

\subsection{Algorithm}

\noindent
{\bf Highlevel Outline}

$\ha$ is managed as a stack. The main  loop computes $A$ and then does a DDFS
anchored on $A$. If an ap is discovered all of $V(A)$ gets deleted from the graph (by definition of anchored DDFS). If a bottleneck $b$ is discovered,
$b=\bd^*(A_i)$, we can merge $\pl^*$s $A_i,\ldots, A_\ell$
(again definition of anchored DDFS).  
So pop the $\hb$ stack to erase the buds above $b$, and add the new
vertices of the $\pl^*$ to $\ha$.
Also merge appropriate bookeeping lists for the new $\pl^*$.
Then  do a dfs from $A_i$, using props,
trying to complete a new $A$-path (i.e., 
search for a base or blossom that has an incident bridge). 
If one is found  start the next iteration. (Do the same thing in the ap case.)
If not, $A_i$ cannot be on any ap. Delete $A_i$
from the graph, and dfs search from $A_{i-1}$.

\b\noindent{\bf Detailed Algorithm}

\subsubsection{Computing $A$}

In the following, each time  $A$ is extended by a new vertex or 
\pls. (which must be a blossom plus its base) $A_i$,
push $V(A_i)$ onto $\ha$ and push $\bd^*(A_i)$ onto $\hb$.

We start with 
$A$ as the sequence $A_1,\ldots, A_i$ ($i=\epsilon$) from the previous search;
if $A$ is empty, initialize it to a free \pls.  $A_1$, and set $i=1$.

Do a dfs from $A_i$, using props to find the next vertex or \pls..
Extend the dfs path $A$ (0 or more times) until  it reaches a vertex or $\pl^*$
$A_\ell$ that either is either a dead end or it has an incident bridge.

If $A_\ell$ has no outgoing props, it cannot be in any ap. Delete $A_\ell$
from $A$ and the graph. Continue the dfs.

If $A_\ell$ has an incident bridge $rg$, do a DDFS anchored on $A$.

\subsubsection{Anchored DDFS}

At any time in the DDFS 
say each DFS path ends at its {\em active vertex}
(called ``center of activity'' in \cite{V}).
We will maintain the invariant that some vertex $\alpha\in V(A)$
is active;
more precisely there is always an index $\epsilon$
such that $\alpha=\bd^*(A_\epsilon)$ 
is active and 
the DDFS has  visited $\bd^*(A_i)$  
iff $i\ge \epsilon$.
Whenever the algorithm moves $\alpha$, it descends to the preceding
vertex of $A$. Thus the invariant is preserved. The invariant implies
that the DDFS correctly implements anchored DDFS (i.e.,
if an ap is  found then $V(A)\con X$;
if not then $X$ is a $\pl^*$
with bud  $\bd^*(A_\epsilon)$ for some $\epsilon$,
and $V(A)\cap X = \bigcup_{i\ge \epsilon} V(A_i)$).
Let $u$ denote the other active vertex (perhaps $u=\alpha$).

The algorithm always records $\epsilon$.
It moves $\alpha$ by decrementing $\epsilon$ by 1.

The rule for how the DDFS moves is 
determined by its state. The 2 possible states are
called {\em Disjoint} (when $u\notin V(A)$) and
{\em Meet} (when $u=\alpha$). (There may be $>1$ move
between states, e.g., the algorithm starts backtracking when
it enters state {\em Meet}.)

\b
\noindent {\em State Disjoint, $u\notin V(A)$.}

If $\l(\alpha)\ge \l(u)$, 
move $\alpha$ to the preceding  vertex of $A$. 
Stay in state Disjoint and make the next move.

If $\l(u)>\l(\alpha)$, 
let $uv$ be a prop from $u$ to a lower level.
(It exists since we're not backtracking.)
If $a[v]=0$, move $u$ to $\bd^*(v)$.
Stay in state Disjoint and make the next move.

If not
(i.e., $a[v]\ne 0$)
 then $v\in V(A)$. 
If $\epsilon>1$ let $b=\hb[\epsilon-1]$, i.e.,  the bud preceding $\alpha$
in $A$. 
If $a[v]\ge a[b]$, $\bd^*(v)$ has been visited, so reject prop $uv$.
Otherwise 
$a[v]< a[b]$, so
$a[v]\le a[\alpha]$.
Move $\alpha$ 0 or more times
until $a[\alpha]\le a[v]$.
Move $u$ to $\alpha$. Now the 2 DFS paths have met
so enter state Meet.

\b
\noindent {\em State Meet, $u=\alpha$.}

Enter backtracking mode. Proceed as usual.
Note that $\alpha$ always remains active during the backtracking.
As before
the backtracking DFS always backs off from a previously visited vertex.
It scans the next prop following the rule for state Disjoint, until
(as usual) it finds a vertex $u\ne \alpha$ at level $\le \l(\alpha)$. 
If $u\notin V(A)$ it goes to state Disjoint.
If $u\in V(A)$, it will reenter state Meet at the new, lower $\alpha$.

\b
\noindent {\em Initializing the DDFS for bridge $rg$.}

Set $\epsilon=\ell$, $\alpha=\bd^*(A_\epsilon)$.
If $a[g]=0$ then set $u=\bd^*(g)$ and begin the DDFS in state Disjoint.
Otherwise $g\in V(A)$. If $a[g]\ge a[\alpha]$ then
$r$ and $g$ are in the same $\pl^*$ so the DDFS terminates.
Otherwise $a[g]<a[\alpha]$, i.e., $g$ is in a $\pl^*$ of $A$
preceding $A_\epsilon$. As in state Disjoint, move $\alpha$ to catch up
with $g$, and enter state Meet.

\subsubsection{Finishing the DDFS}

If an ap is discovered, delete it from the graph, and (as in \cite{V}),
repeatedly delete all vertices and $\pl^*$'s with no predecessors.
This deletes
$V(A)$, by definition of anchored DDFS. 

If a bottleneck $b$ is discovered,
$b=\alpha=\bd^*(A_\epsilon)$.
Merge $\pl^*$s $A_\epsilon,\ldots, A_\ell$. In the data structure  
pop the $\hb$ stack to erase the buds above $b$, and add the new
vertices of the $\pl^*$ to $\ha$.
Also merge appropriate bookeeping lists into those
for the new $\pl^*$.

Now in either case start the next iteration.
 
\subsection{Time Bound}
An anchored DDFS is a special case of DDFS.
So as in the Micali-Vazirani algorithm, all the
anchored DDFSs use total  time $O(m)$ plus the time for
set merging. It is easy to see we use
$O(1)$ time to identify the \bds. of a vertex in $A$.
Thus all anchored DDFSs use total time $O(m)$.

Beyond DDFS the algorithm performs a number of dfs's to extend
$A$. An edge of the graph gets scanned at most once in all these
searches. Hence they use total time $O(m)$.

\end{document}